# Heat Management in Thermoelectric Power Generators

M. Zebarjadi,[1,2,*]

[1] Department of Mechanical and Aerospace Engineering, Rutgers University, Piscataway, New Jersey, 08854, USA

[2] Institute of Advanced Materials, Devices, and Nanotechnology, Rutgers University, Piscataway, NJ, 08854, USA

* Author to whom correspondence should be addressed.  Electronic mail: m.zebarjadi@rutgers.edu.

**ABSTRACT**

Thermoelectric power generators are used to convert heat into electricity. Like any other heat engine, the performance of a thermoelectric generator increases as the temperature difference on the sides increases. It is generally assumed that as more heat is forced through the thermoelectric legs, their performance increases. Therefore, insulations are typically used to minimize the heat losses and to confine the heat transport through the thermoelectric legs. In this paper we show that to some extend it is beneficial to purposely open heat loss channels in order to establish a larger temperature gradient and therefore to increase the overall efficiency and achieve larger electric power output.  We define a modified Biot number (Bi) as an indicator of requirements for sidewall insulation. We show that if Bi<1, cooling from sidewalls increases the efficiency, while for Bi>1, it lowers the conversion efficiency.

**Introduction**

Thermoelectric power generators are proposed for many applications such as waste heat recovery (as topping cycles)[1–4], automobile industry, and solar thermoelectric power generators[5–7]. It has been shown that the performance of the thermoelectric power generators is an increasing function of its material figure of merit, $ZT = \frac{\sigma S^2 T}{\kappa}$, where in $\sigma$ is the electrical conductivity, S is the Seebeck coefficient, T is the operating temperature and $\kappa$ is the thermal conducitity.[2] Typical commercial thermoelectric devices are made out of $Bi_2Te_3$/$Sb_2Te_3$ for room temperature applications and PbTe for high temperature applications and they have a ZT of around 1. These modules are made out of many p-n legs, which are placed thermally in parallel and electrically in series.  As an example, HZ-14 model developed by Hi-Z Company, has 6.27cm by 6.27cm ceramic area with 49 p-n pairs of bismuth telluride based semiconductors and a thickness of about 5mm.[8] The module provides 25W (5% efficiency) output for a temperature difference of 300°C. This is equivalent of $\nabla T$ =60°C/mm which is quite large. According to the Goldsmid analysis[9], the ideal efficiency of the thermoelectric module used in this example (i.e. for a ZT~1 and the temperature difference of 300°C) is about 10%. Clearly, there is a factor-of-two difference between ideal and practical efficiency, which could be attributed to thermal/electrical contact resistances, and non-ideal heat managements (operational conditions). Both issues have been largely studied in the literature in the past. In this work, we would like to address the heat management issue. It has been shown by several authors, that heat loss through sidewalls of thermoelectric legs, result in lowering of the thermoelectric efficiency.[10–12] The argument here is simple. Opening of heat loss channels, results in less conversion of heat to electricity and therefore lowering the thermoelectric efficiency, a direct conclusion. The purpose of this article is to point to practical issues resulting in cases, where in opening of heat loss channels, improves the efficiency!

**Methods**

Consider a single n-p thermoelectric module schematically shown in Fig. 1. We assume constant materials properties in each leg. $S_{n/p}, \rho_{n/p}, \kappa_{n/p}$ are the Seebeck coefficient, the electrical resistivity and the thermal conductivity of the n/p legs respectively. The n-p legs are connected electrically in series and thermally in parallel. Therefore the electrical resistance, R, and the thermal conductance, K, of a single n-p pair, ignoring the metallic connections and ignoring the interfacial resistances, could be written as: $R = \frac{\rho_p l_p}{A_p} + \frac{\rho_n l_n}{A_n}$ and $K = \frac{\kappa_p A_p}{l_p} + \frac{\kappa_n A_n}{l_n}$ (this is the total thermal conductance and has the contribution of electrons and phonons). The Seebeck coefficient of the n-p pair is $\alpha = (S_p - S_n)$ and finally, figure of merit is defined as $Z = \frac{\alpha^2}{RK}$.

In a pioneer work, Goldsmid[13] developed an analytical model, wherein he assumed one dimensional transport within the thermoelectric legs, neglecting convective heat loss from the perimeter (perfect insulation) and also neglecting contacts. He applied constant temperature boundary conditions ($T = T_H$ at hot side and $T = T_C$ at cold heat sink), and proved that the maximum achievable efficiency (for optimum external load) could be written in terms of n-p figure of merit (Z) and the temperature difference ($\Delta T = T_H - T_C$) as:

$$\eta = \frac{(\gamma-1)\Delta T}{(\gamma+1)T_H - \Delta T} \tag{1}$$

$$\gamma = \sqrt{1 + ZT_M} \tag{1a}$$

$$T_M = \frac{T_H + T_C}{2} \tag{1b}$$

It is clear from Eq. 1 that larger Z values and larger temperature differences result in larger efficiencies. Consequently, the natural tendency is to 1) insulate thermoelectric legs to minimize heat loss and run the module as close as possible to the ideal conditions (perfect insulation) and 2) impose a large temperature gradient by connecting one end to a heat source with high heat fluxes and large temperatures, and cooling down the other end by cold air/water fluxes. There is no doubt that such approach is correct if we assume temperature of the hot/cold ends (of the thermoelectric legs) are exactly the same as the hot heat source/cold heat sink. In practice this is not the case. There is always a temperature drop at the interface of the hot/cold side. If the cold end is cooled down by a fluid flux at temperature $T_F$, the temperature at the cold end of the thermoelectric leg is not equal to and is larger than the fluid temperature $T_C > T_F$. The correct boundary condition in this case, is matching the heat flux at the cold side; to the convective heat transfer flux from the thermoelectric leg to the air ($Q' = h_F(T_C - T_F)$). Only when the heat transfer coefficient of the fluid goes to infinity ($h_F \to \infty$), $T_C = T_F$ and constant temperature boundary conditions can be used. In many cases, cooling of the cold side is too expensive and thermoelectric modules are simply attached to a heat source and the cold end is cooled by natural convection for which, $h_F$ is only about 1W/m²K. In the case of forced air convection (using a fan), $h_F$ can increase to about 100W/ m²K. Water-cooling is more expensive but it can increase the heat transfer coefficient to quite large values (10-1000 W/m²K) and to even larger values when force water-cooling is used.

In cases wherein poor cooling is performed, a natural consequence of sending a large heat flux to the thermoelectric modules, is over heating of the module and consequently establishment of much smaller temperature differences ($\Delta T < T_{hot-source} - T_{heat-sink/fluid}$) which therefore lowers the overall efficiency. In such cases, it might be possible to increase the temperature difference simply by opening heat loss channels at sidewalls. That is to remove the insulation layers and allow a larger surface area to be in contact with the cooling source to establish a larger temperature difference along the leg. The main question is what is the best operating conditions in which efficiency is large enough with cheaper cooling options.

To answer this question, we develop a more realistic model considering convective heat transfer loss from the perimeter and considering a more realistic convective boundary condition at the cold side. Only if the heat transfer coefficient goes to infinity, the boundary condition at the cold side is $T = T_C$. The geometry and the detailed derivation of the heat conduction equation are shown in Fig.1.

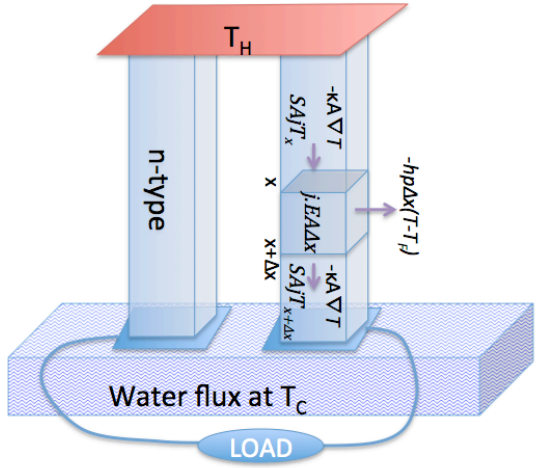

**Figure. 1.** Schematic diagram representing a single n-p pair thermoelectric module. The module is in well contact with a heat source and from the other end is cooled down by a liquid flux. Energy balance over a slab of p leg is shown and heat loss through the sidewalls are accounted for using convective heat transfer.

Here to simplify we only write the equation for the p leg and therefore drop the sub-index p for materials/geometrical properties. The results could be simply extended to include both types by defining total thermal/electrical resistances of the n-p leg. The results of writing the heat balance equation, as shown in figure 1, and as discussed in our previous publication[14], is a heat conduction equation for electrons and phonons combined.

$$\frac{d^2T}{dx^2} - \frac{hP}{\kappa A}(T - T_a) + \frac{\rho j^2}{\kappa} = 0 \qquad (2)$$

and the heat flux is defined as:

$$Q'_x = -\kappa \frac{dT}{dx}\bigg|_x + ST_x j \qquad (3)$$

$P$ is the leg perimeter, A is the cross section of the leg, $T_a$ is the ambient temperature, $\kappa$ is the total thermal conductivity due to electrons and phonons, and $j$ is the electric current flux. The main approximation here is that temperature is constant in the y-z plane. Note that this assumption is used very often in modeling of fins.[15,16] Also note that in Eq. 2, we are neglecting the Thomson effect to simplify the solutions. In other words, we assume that the Seebeck coefficient is not changing with temperature ($\frac{dS}{dT} = 0$). Full solutions with inclusion of Thomson terms, are too lengthy and it is hard to extract information from such complicated equations.[12,17]

To further simplify the solutions, we define a set of dimensionless parameters and we also change the reference for measuring the temperature:

$$T_I = \frac{\rho j^2}{\kappa \omega^2} = \frac{J^2}{Z(\omega l)^2} \tag{4a}$$

$$T_\omega = T_a + T_I \; ; \; T_{H\omega} = T_H - T_\omega \; ; \; T_{aI} = T_a - T_I \; ; T_{x\omega} = T_x - T_\omega \tag{4b}$$

$$J = \frac{SI}{K} \tag{4c}$$

$$Bi = \frac{h_F l}{\kappa} = \frac{h_F A}{K} \tag{4d}$$

$$\omega l = l\sqrt{\frac{hP}{\kappa A}} = \sqrt{\frac{hPl}{K}} \tag{4e}$$

$J$ is a dimensionless current, $Bi$ is similar to the Biot number, defined at the cold end with $l$ being the length of the thermoelectric leg and $\kappa$, the thermal conductivity of the TE module. $Bi$ is representative of the effectiveness of cooling at the cold side. Larger $Bi$ values correspond to larger $h_F$ values and therefore better cooling at the cold end. $\omega l$ is another dimensionless parameter, which is known as the fin parameter. In this analysis, it is the representative of heat loss through sidewalls and increases as $h$ (the heat transfer coefficient through the sidewalls) increases.

The solution of Eq. 2, for fixed temperature boundary condition ($T_0 = T_H$) at the hot side and convective boundary conditions at the cold side ($Q'_l = -\kappa \frac{dT}{dx}|_{x=l} + ST_l j = h_F(T_l - T_a)$) is

$$T_{x\omega} = e^{-\omega x} T_{H\omega} + \frac{e^{-\omega l}(-Bi + J + \omega l) T_{H\omega} - T_I \, Bi + J \, T_\omega}{\omega l \, Cosh(\omega l) + (Bi - J) Sinh(\omega l)} Sinh(\omega x) \tag{5}$$

Note that here to simplify the problem; we assumed that the temperature of the cooling fluid ($T_F$) is the same as the ambient temperature ($T_a$). The difference between the cooling of the cold side and the heat loss through the sidewalls is shown with the two different heat transfer coefficients $h$ and $h_F$.

Using Eq. 3, the heat rate at the hot side (x=0) is:

$$Q_H = SIT_H + K\omega l \frac{Bi \ T_I - J \, T_\omega + (Bi - J) T_{H\omega} \cosh(\omega l) + \omega l \, T_{H\omega} \sinh(\omega l)}{\omega l \cosh(\omega l) + (Bi - J) \sinh(\omega l)} \tag{6}$$

It can be shown that the useful work done on the external load is:

$$W = Q_H - Q_C - Q_{loss} = (S\Delta T - RI)I \tag{7a}$$

$$W = \frac{JK}{Z} \frac{-\omega l \, ZT_{H\omega} + \omega l (ZT_{H\omega} - J) \cosh[\omega l] + (-Bi \, J + J^2 - ZT_H J + Bi \, ZT_a) \sinh[\omega l]}{\omega l \cosh[\omega l] + (Bi - J) \sinh[\omega l]} \tag{7b}$$

Finally the efficiency is

$$\eta = \frac{W}{Q_H} = \frac{2 J \omega l \, (ZT_{H\omega} + (J - ZT_{H\omega}) \cosh[\omega l] + [(Bi - J)(J - ZT_H) + Bi \, ZT_a] \frac{\sinh[\omega l]}{\omega l})}{e^{-\omega l}(Bi - J - \omega l)(J \, ZT_H - \omega l \, ZT_{H\omega}) - e^{\omega l}(Bi - J + \omega l)(J \, ZT_H + \omega l \, ZT_{H\omega}) + \frac{2J}{\omega l}(J^2 - Bi \, J + (\omega l)^2 \, ZT_a)} \tag{8}$$

**Results**

To look at a practical device performance, we take dimensions from a typical thermoelectric module. For example those of the HZ-14 discussed earlier in the introduction and is summarized in caption of Fig.2.

Fig. 2a shows the temperature distribution along thermoelectric legs, when the hot side is kept at 600K, and the cold side is cooled down by a coolant fluid of temperature $T_F = 300K$ and heat transfer coefficient of $h_F = 100 \frac{W}{m^2K}$. Convective heat loss from sidewalls is also considered, assuming that the walls are in contact with a fluid at ambient temperature, which is taken to be the same as the coolant temperature $T_a = 300K$, but with a different heat transfer coefficient ($h$). Graphs for different applied currents and two different $h$ values are plotted. To prevent any prior assumption on the Seebeck coefficient, instead of current we use SI (Seebeck times current), which could be tuned by tuning the electrical current through the external load. As current increases, the temperature difference on the sides of the TE leg decreases as a result of Joule heating. This is shown in the inset of Fig. 1a. The input heat pumped at the hot side is plotted in Fig 2b and increases as more current is passed (as current carries more heat into the leg). Finally there is an optimum work and efficiency versus current (Figs. 2c and 2d). Figure 2 is plotted for two different sidewall heat transfer coefficients. Even at this relatively large coolant heat transfer coefficient ($h_F = 100 \frac{W}{m^2K}$), we can clearly see that increasing the heat loss through the sidewalls (i.e. increasing $h$ from 50 to $100\frac{W}{m^2K}$), results in establishment of larger temperature differences and therefore larger efficiencies as shown in Fig. 2d.

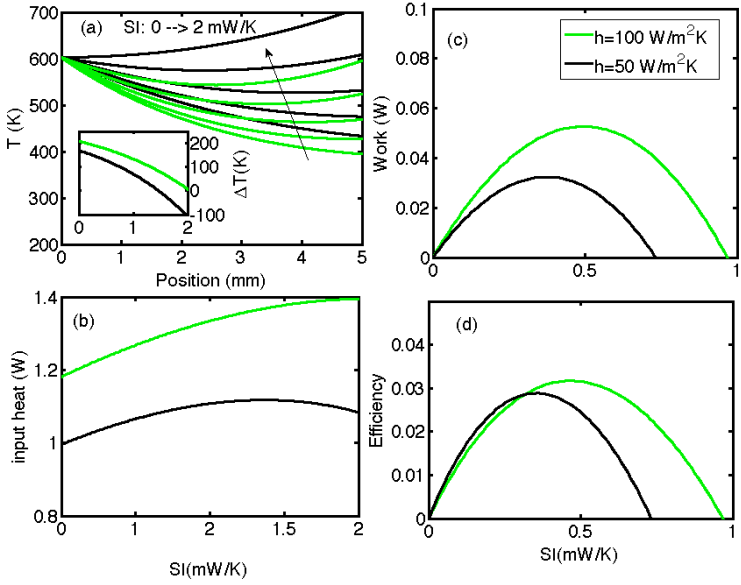

**Figure. 2.** Results for a single p-n pair, the dimensions are $4 \times 4 \times 5 \ mm^3$,

$T_H = 600K, T_a = T_F = 300K, h_F = 100 \frac{W}{m^2 K}, ZT_M = 1, \kappa = 1 \frac{W}{mK}$. Black lines are plotted for $h = 50 \frac{W}{m^2 K}$ and green lines for $h = 100 \frac{W}{m^2 K}$. a) Temperature distribution along the thermoelectric leg for different SI values which is increased from 0 to $2 \frac{mW}{K}$ with steps of $0.5 \frac{mW}{K}$ and using Eq. 5. Inset: $\Delta T = T_0 - T_l$ as a function of SI (in $\frac{mW}{K}$ units). b) Input heat current (heat current at the hot side, $Q_H$, Eq. 6) as a function of SI, c) Useful output work (Eq. 7) as a function of SI and d) Efficiency of heat to electricity conversion (Eq.8) as a function of SI.

We know at the limit of $h_F \rightarrow \infty$, the results will be completely different and larger $h$ values, result in more heat loss and less efficiency. So the next natural question is that how large is large enough to be considered as the limit of infinity and when we will switch from gaining by opening heat loss channels to loosing by opening such channels. To answer this question, we note that for a fixed set of temperatures, efficiency is a function of 4 different parameters namely Z, J, $\omega l$, and $Bi$ (see Eq. 8).

Among these parameters, J is irrelevant in the sense that it is tunable and for a given thermoelectric module, we can optimize the efficiency versus this parameter. Z parameter could be fixed, as we know good commercial thermoelectric materials possess $ZT_M$ values close to 1 and in the research labs this value is close to 2. The remaining two parameters, $\omega l$ and $Bi$, are then the crucial parameters in our discussion. While $\omega l$ represents the amount of heat loss from the sidewalls, $Bi$, represents the effectiveness of the cooling system used at the cold end. To find the transition point, we fixed the temperatures and the figure of merit. Then, for a given $\omega l$ and $Bi$, we optimized the efficiency versus J. The resulted optimum efficiency is plotted in Fig.3. For the set of chosen parameters, the ideal efficiency, according to the Goldsmid analysis, Eq. 1, is about 10%. Fig. 3 shows that this ideal efficiency is only achievable for $Bi = \frac{h_F l}{\kappa}$ values larger than 8. For a thermal conductivity of 1W/mK, and length of 5mm, this corresponds to $h_F > 1600 \frac{W}{m^2 K}$. In this regime, as expected, any heat loss (increase of $\omega l$) results in lowering of the efficiency. The transition happens at around $Bi \sim 1$. Below this value, the efficiency increases as heat loss through sidewalls increases (when $\omega l$ increases), which is counter-intuitive and is happening due to the consequent establishment of a larger temperature difference. The transition point could be more easily identified from Fig. 3b. In this figure the derivative of the efficiency with respect to $\omega l$ is plotted. This derivative is positive for small Bi values and is negative for larger ones. The transition happens around $Bi \sim 1$. The absolute values of efficiency are sensitive to the chosen temperatures (of the hot side and of the ambient/fluid) and the Z parameter. However, the overall shape of this graph (Fig. 3) has only minor sensitivity to any of these parameters and therefore the $Bi \sim 1$ value could be taken as the transition value independent of the Z and the temperature values. In fact, the overall result of increasing Z parameter and temperature difference is to shift the transition point slightly to larger values. Which means for larger ZT values, even larger $h_F$ values are required for efficient heat conversion. The studied case of Fig.2 was corresponding to $Bi = 0.5$ which is clearly in the regime wherein heat loss through sidewalls is beneficial.

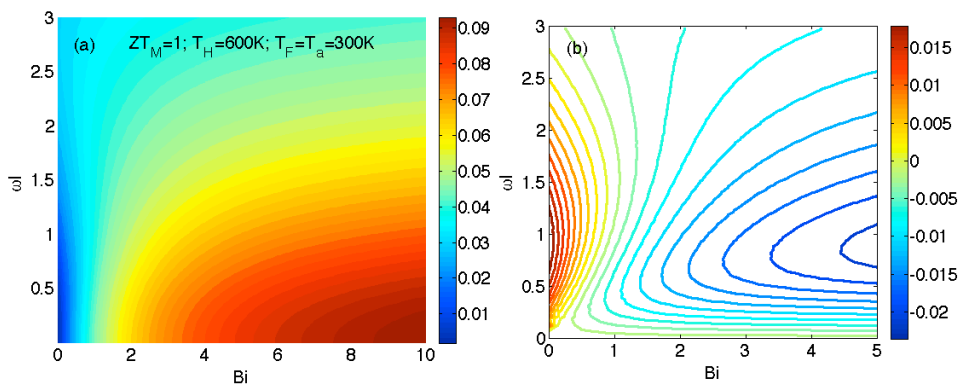

**Figure 3.** a) Optimum efficiency versus $Bi$, and $\omega l$, plotted for $T_H = 600K, T_F = T_a = 300K, and\ ZT_M = 1$. Ideal efficiency (from Eq. 1) for the set of chosen parameters is 10%. b) The derivative of efficiency with respect to $\omega l$ using the same parameters as in part a)

## Conclusions

In summary, we developed a fin model for thermoelectric power generators, including the convective heat transfer coefficient from the sidewalls. We determined that in addition to the Z parameter and the temperatures, the device efficiency is also a function of two dimensionless parameters. One is ($Bi = \frac{h_F l}{\kappa}$, Biot number) which represents the effectiveness of the cooling system used through $h_F$ value and the other is ($\omega l = l\sqrt{\frac{hP}{\kappa A}}$, fin parameter) which represents the heat loss through side walls. We showed that if $Bi < 1$, that is when poor cooling is used at the cold side, opening heat loss channels through the sidewalls, increases the overall efficiency of the thermoelectric module as a result of formation of a larger temperature difference over the thermoelectric legs. In the opposite regime, when $Bi > 1$, we recovered the normal predicted behavior of thermoelectrics, where in extra heat loss result in lowing of the heat to electricity conversion efficiency.

## Acknowledgements

MZ would like to acknowledge K. Esfarjani for his helpful feedback on the manuscript. This work is supported by the Air Force young investigator award, grant number FA9550-14-1-0316.